\documentclass[twocolumn,showpacs,superscriptaddress,amssymb,10pt]{revtex4}

\usepackage{graphicx}
\usepackage{dcolumn}
\usepackage{bm}
\usepackage{epsfig}
\usepackage{color}
\usepackage{longtable}

\def\ket#1{ $ \left\vert  #1   \right\rangle $ }
\def\ketm#1{  \left\vert  #1   \right\rangle   }

\def\sprm#1#2{  \left\langle #1 \left\vert \right. #2 \right\rangle   }

\def\mem#1#2#3{  \left\langle #1 \left\vert  #2 \right\vert #3 \right\rangle   }
\def\rmem#1#2#3{  \left\langle #1 \left\vert \left\vert  #2
                  \right\vert \right\vert #3 \right\rangle   }

\def\sixjm#1#2#3#4#5#6{  \left\{ \begin{array}{ccc}
                                               #1 & #2 & #3  \\
                                               #4 & #5 & #6
                     \end{array} \right\}   }
\def\asr#1{#1}        

\begin{document}

\preprint{}

%
%
%
%
\title{Hyperfine--induced effects on the linear polarization of\\[0.1 cm] the K$\alpha_1$ emission from helium-like ions}

%
%
%
%

\author{Andrey Surzhykov}
\affiliation{Physikalisches Institut, Ruprecht--Karls--Universit{\"a}t Heidelberg, D--69120 Heidelberg, Germany}
\affiliation{GSI Helmholtzzentrum f\"ur Schwerionenforschung, D--64291 Darmstadt, Germany}
\affiliation{Helmholtz--Institut Jena, D--07743 Jena, Germany}

\author{Yuri Litvinov}
\affiliation{GSI Helmholtzzentrum f\"ur Schwerionenforschung, D--64291 Darmstadt, Germany}

\author{Thomas~St{\"o}hlker}
\affiliation{GSI Helmholtzzentrum f\"ur Schwerionenforschung, D--64291 Darmstadt, Germany}
\affiliation{Helmholtz--Institut Jena, D--07743 Jena, Germany}
\affiliation{Institut f\"ur Optik und Quantenelektronik, Friedrich--Schiller--Universit\"at Jena, D--07743 Jena, Germany}

\author{Stephan Fritzsche}
\affiliation{Helmholtz--Institut Jena, D--07743 Jena, Germany}
\affiliation{Theoretisch--Physikalisches Institut, Friedrich--Schiller--Universit\"at Jena, D--07743 Jena, Germany}

\date{\today \\[0.3cm]}

%
%
%
%
\begin{abstract}
The linear polarization of the characteristic photon emission from few--electron ions is studied for its sensitivity with regard to the nuclear spin and magnetic moment of the ions. Special attention is paid, in particular, to the K$\alpha_1$ ($1s 2p_{3/2} \;\, ^{1,3}P_{1,2} \to 1s^2 \;\, ^1S_0$) decay of selected helium--like ions following the radiative electron capture into initially hydrogen--like species. Based on the density matrix theory, a unified description is developed that includes both, the many--electron and hyperfine interactions as well as the multipole--mixing effects arising from the expansion of the radiation field. It is shown that the polarization of the K$\alpha_1$ line can be significantly affected by the mutipole mixing between the leading $M2$ and hyperfine--induced $E1$ components of $1s2p \;\, {}^3P_2, \, F_i \,\to\, 1s^2 \;\, {}^1S_0, \, F_f$ transitions. This $E1$--$M2$ mixing strongly depends on the nuclear properties of the considered isotopes and can be addressed experimentally at existing heavy--ion storage rings.
\end{abstract}

\maketitle

%
%
%
%
\section{Introduction}

The application of atomic spectroscopy techniques for studying nuclear properties has a very long tradition that goes back to the early days of modern physics. Over the years, a large number of experiments have been carried out in order to determine, for example, isotope shifts of optical as well as x--ray transitions along various chains of isotopes, or to study the hyperfine structure of atomic levels. When compared to predictions from atomic theory, these measurements helped reveal important information about the spins, moments and charge radii of nuclear ground and isomeric states (see e.g.~\cite{BiC95,Ang09}). Moreover, the last decades have witnessed a significant progress in developing optical laser spectroscopy techniques that allow probing of the structure of stable and especially radioactive nuclei. Combined with modern accelerator and storage ring facilities, these techniques allow very accurate investigations of short--living exotic isotopes which are available only in small quantities~\cite{KlN03,Klu10,Charlwood:10,Cocolios:11,Blaum13}.

While most atomic--spectroscopy studies on the shape and electro--magnetic moments of the nuclei usually deal with (shifts in the) transition energies or lifetimes of the excited levels, less attention has been paid so far to the analysis of the angular and polarization properties of bound--state atomic (or ionic) transitions. Owing to the hyperfine coupling, however, these properties might be sensitive also to the spin $I$ as well as the magnetic $\mu$ and the quadrupole $Q$ moments of the nucleus. For electron--impact excited helium-like Sc$^{19+}$ ions, for example, the linear polarization of the characteristic K$\alpha$ lines was found to be strongly affected by the hyperfine interaction and, hence, the nuclear properties~\cite{Hen90,DuG94,BeI06}. Recent advances in x--ray detector techniques~\cite{LeD97,Sch98,Sof03,PrS05,Tas06,Spi08,Web10,Mar11,Mar12} suggest that high--precision polarization measurements of hyperfine transitions are possible in the near future and may provide a very promising route for probing the properties of stable and radioactive isotopes. When performed at ion storage rings, such measurements will become feasible not only for low-- and medium--$Z$ elements but also for heavy, few--electron ions \cite{Fritzsche:05,FAIR-report,Sto13}.

In this contribution we present a theoretical study of the characteristic photon emission from few--electron ions with non--zero nuclear spin, $I \ne 0$. To explore the linear polarization of these transitions, a general formalism is laid down that accounts for both, the many--electron correlation effects as well as the higher--order multipole components of the radiation field, and that includes the hyperfine--induced channels. While this formalism can be applied to \textit{all} ions, independent of their particular shell structure, special attention is paid to the $1s 2p_{3/2} \;\, ^{1,3}P_{1,2} \to 1s^2 \;\, ^1S_0$ (referred to as K$\alpha_1$) decay of heavy, helium--like ions. For these ions, the angular and polarization properties of the K$\alpha_1$ lines can be investigated experimentally at storage ring facilities by applying the radiative electron capture (REC) of initially hydrogen--like ions \cite{EiS07, SuJ06a} to form the $1s 2p_{3/2} \;\, J=1,2$ excited states. For the subsequent K$\alpha_1$ emission of the ions into their $1s^2 \;\, ^1S_0$ ground state, a remarkable shift is shown especially for the linear polarization owing to the interference between the leading $M2$ and hyperfine--induced $E1$ decay channels of the $1s 2p \;\, ^{3}P_{2} \, \to 1s^2 \;\, ^1S_0$ fine--structure transition. Moreover, since the magnitude of the $E1$--$M2$ multipole mixing strongly depends on the (nuclear) spin $I$ and magnetic dipole moment $\mu_I$ of the corresponding isotope, we analyze and argue below how the x--ray polarimetry of the characteristic K$\alpha_1$ emission can be utilized as a tool for studying these nuclear properties.

%
%
%
%
\section{Theoretical background}

During the last decades, various theoretical studies have been carried out in order to analyze the angular distribution and polarization of the characteristic radiation following the excitation of atoms and ions \cite{BeI06,EiS07,SuJ06a,InD89,ReC93,Blu81,BaG00,SuF02,FrS09}. Most naturally, these investigations are performed within the framework of the density matrix theory in which the population of an (excited) atomic states is described in terms of the so--called statistical tensors $\rho_{kq}$, and which are constructed in order to transform like spherical harmonics of rank $k$ under a rotation of the coordinates. Of course, the particular form of these tensors depends not only on the nuclear and electronic shell structure but also on the way how the excited states were generated in an experiment. For the radiative capture of unpolarized electrons into a hyperfine level $\ketm{\alpha_i F_i}$, for example, these tensors can be written as \cite{BaG00, FrS05}:
\begin{eqnarray}
   \label{eq_stat_tensors_F}
   \rho_{kq}(\alpha_i F_i) & = &
   \delta_{q 0 } \, \frac{2F_i+1}{2I+1} \, (-1)^{I + J_i + F_i + k}   \nonumber \\[0.2cm]
   & & \times \; \sixjm{F_i}{F_i}{k}{J_i}{J_i}{I} \, \rho_{k0}(\alpha_i J_i) \, ,
\end{eqnarray}
where $I$ and $J_i$ are the nuclear and total electron angular momenta, $\bm{F}_i = \bm{I} + \bm{J}_i$, and where $\alpha_{i}$ denote all additional quantum numbers as required to specify the state uniquely. In Eq.~(\ref{eq_stat_tensors_F}), we have assumed moreover that the---relatively weak---hyperfine interaction does not affect the recombination process and, hence, that the sublevel population of the $\ketm{\alpha_i F_i}$ state is defined solely by the $I-J$ coupling. In this approximation, the $\rho_{kq}(\alpha_i F_i)$ is proportional to the corresponding statistical tensor of the electronic state, $\ketm{\alpha_i J_i}$, and which was worked out and evaluated before within the multiconfiguration Dirac--Fock method~\cite{SuJ06}.

\asr{In order to derive the statistical tensors $\rho_{kq}(\alpha_i F_i)$ in expression (\ref{eq_stat_tensors_F}) we have adopted the quantization ($z$--) axis of the overall system along the momentum of the incoming electron as \textit{seen} in the rest frame of the ion. This choice of the quantization axis is convenient to explore the sublevel population of the residual ion following the electron capture. Namely, the normalization of the statistical tensors (\ref{eq_stat_tensors_F}) by means of the \textit{zero}--rank tensor gives rise to the \textit{alignment} parameters ${\mathcal A}_k(\alpha_i F_i) = {\rho_{k0}(\alpha_i F_i)}/{\rho_{00}(\alpha_i F_i)}$, and which are related also to the partial cross sections $\sigma(\alpha_i F_i M_i)$ for the capture of an electron into particular hyperfine sublevels \ket{\alpha_i F_i M_i}:}
\begin{eqnarray}
   \label{eq_alignment_cross_section_relation}
   {\mathcal A}_k(\alpha_i F_i) &=& \frac{\sqrt{2F_i + 1}}{\sum\limits_{M_i} \sigma(\alpha_i F_i M_i)}
   \nonumber \\
   && \hspace{-2.0cm} \times \sum\limits_{M_i} (-1)^{F_i - M_i} \, \sprm{F_i M_i \, F_i - M_i}{k 0} \, \sigma(\alpha_i F_i M_i) \, .
\end{eqnarray}
\asr{Apart from the capture of a free (or quasi--free) electron directly into a given level $\ketm{\alpha_i F_i}$, its sublevels can be populated also by cascade decay from some higher--lying states. For relativistic collisions of heavy ions with low--$Z$ targets, such a cascade feeding may significantly affect the alignment of excited ionic states \cite{EiS07}. To take the cascade effects into account, in the present work we also included the radiative electron capture into the $1s \, nl$ levels with $n \le 6$, together with their subsequent decay, and solved the corresponding system of rate equations
(see Ref. [35] for further details). Based on this analysis, we found that the  alignment parameters ${\mathcal A}_k(\alpha_i F_i)$ of the levels of interest are reduced due to cascade effects by about 20~\% for low--energy electron--ion collisions and almost by a factor of two for high collision energies. These cascade contributions are taken into account in the computations below.}

The alignment parameters ${\mathcal A}_k(\alpha_i F_i)$ can be utilized in order to express the (normalized to unity) angular distribution of the photons emitted in the subsequent radiative decay $\ketm{\alpha_i F_i} \,\to\, \ketm{\alpha_f F_f} \,+\, \gamma \,$ into some lower--lying hyperfine level $\ketm{\alpha_f F_f}$ of the ions \cite{SuJ06a,Blu81,BaG00}:
\begin{eqnarray}
   \label{eq_angular_distribution}
   W_{if}(\theta) = \frac{1}{4 \pi}
   \left( 1 + \sum\limits_{k = 2, 4,...} \mathcal{A}_{k}(\alpha_i F_i) \, f_k \, P_k(\cos\theta) \right)  .
\end{eqnarray}
\asr{Since the excited ionic states $\ketm{\alpha_i F_i}$ produced by the capture of unpolarized electrons posses axial symmetry, this distribution depends solely on the polar angle $\theta$ of the decay photon momentum with respect to the quantization axis (i.e. direction of the incident electrons).}

\asr{Beside the anisotropic angular distribution (\ref{eq_angular_distribution}), the alignment of excited ionic states results also in a non--vanishing linear polarization of the characteristic x--ray photons. In atomic physics, this polarization is usually specified by its \textit{degree} \cite{BaG00,BeL71,PeS58}:}
\begin{equation}
   \label{eq_degree_polarization_definition}
   P_{if}(\theta) = \frac{I_{\parallel}(\theta) - I_{\perp}(\theta)}{I_{\parallel}(\theta) + I_{\perp}(\theta)} \, ,
\end{equation}
\asr{where $I_{\parallel}$ and $I_{\perp}$ denote the intensities of light, that is linearly polarized in parallel and perpendicular with respect to the reaction ($xz$--) plane as spanned by the directions of the emerged light and incident electrons. One can see from Eq.~(\ref{eq_degree_polarization_definition}), that the degree $P_{if}$ is also a function of the angle $\theta$ between these two directions. The explicit form of such an angular dependence can be derived most naturally within the framework of the density matrix theory (see e.g. Refs.~\cite{Blu81,BaG00}) and reads:}
\begin{eqnarray}
   \label{eq_linear_polarization}
   P_{if}(\theta) =
   \frac{\sum\limits_{k = 2, 4,...} \sqrt{\frac{16 \pi}{2k + 1}} \,
         \mathcal{A}_{k}(\alpha_i F_i) \, g_k \, Y_{k 2}(\theta, 0)
       }{1 + \sum\limits_{k = 2, 4,...} \mathcal{A}_{k}(\alpha_i F_i) \, f_k \, P_k(\cos\theta)} \, ,
\end{eqnarray}
\asr{where, again, a set of alignment parameters $\mathcal{A}_{k}$ characterizes the population of the excited ionic state.}

As seen from Eqs.~(\ref{eq_angular_distribution}) and (\ref{eq_linear_polarization}), the angular and polarization properties of the characteristic radiation depend, apart from the alignment parameters, on the two functions $f_k \equiv f_k(\alpha_i F_i, \alpha_f F_f)$ and $g_k \equiv g_k(\alpha_i F_i, \alpha_f F_f)$, which are independent of the capture process but merely reflects the electronic shell structure of the ions and its coupling to the nuclear spin. Using the (standard) multipole expansion of the electron--photon interaction operator \cite{Ros57,Gra74}, we can write these functions as:
\begin{widetext}
\begin{eqnarray}
\label{eq_f_structure}
   f_k(\alpha_i F_i, \alpha_f F_f) &=&
   \frac{\sqrt{2 F_i + 1}}{2} \, \sum\limits_{L p \, L' p'}
   i^{L' + p' - L - p} \, [L, L']^{1/2} \: (-1)^{F_f + F_i + k + 1} \, \sprm{L 1 L' -1}{k 0} \,
   \sixjm{L}{L'}{k}{F_i}{F_i}{F_f}
   \nonumber \\[0.1cm]
   && \times  \, \left( 1 + (-1)^{L + p + L' + p' - k} \right) \:
   \rmem{\alpha_i F_i}{H_\gamma(p' L')}{\alpha_f F_f} \,
   \rmem{\alpha_i F_i}{H_\gamma(p L)}{\alpha_f F_f}^*
   \nonumber \\[0.1cm]
   && \times \,
   \left[ \sum\limits_{L p} \left|\rmem{\alpha_i F_i}{H_\gamma(p L)}{\alpha_f F_f} \right|^2 \right]^{-1} \ ,
   \\[0.2cm]
\label{eq_g_structure}
   g_k(\alpha_i F_i, \alpha_f F_f) & = &
   \frac{\sqrt{2 F_i + 1}}{2} \, \sum\limits_{L p \, L' p'} \,
   i^{L' + p' - L - p} \, [L, L']^{1/2} \, (-1)^{F_f + F_i + k + 1 + p'} \,
   \sprm{L 1 L' 1}{k 2} \, \sixjm{L}{L'}{k}{F_i}{F_i}{F_f}  \nonumber \\[0.1cm]
   && \times \,
   \rmem{\alpha_i F_i}{H_\gamma(p' L')}{\alpha_f F_f} \,
   \rmem{\alpha_i F_i}{H_\gamma(p L)}{\alpha_f F_f}^*\nonumber \\[0.1cm]
   && \times \,
   \left[ \sum\limits_{L p} \left|\rmem{\alpha_i F_i}{H_\gamma(p L)}{\alpha_f F_f} \right|^2 \right]^{-1} \, .
\end{eqnarray}
\end{widetext}
Here, the notation $[L] = 2L + 1$ is introduced, and $\rmem{\alpha_i F_i}{H_\gamma(p L)}{\alpha_f F_f}$ denotes the \textit{reduced} matrix element for the magnetic ($p = 0$) or electric ($p = 1$) radiative transition of the order $L$.

In practice, only a few multipole components $(p L)$ are typically allowed in the summation in Eqs.~(\ref{eq_f_structure})--(\ref{eq_g_structure}) owing to the parity and angular--momentum selection rules. Nevertheless, the interference between these components can significantly affect the structure functions $f_k$ and $g_k$ and, hence, the angular and polarization properties of the characteristic radiation.

Equations (\ref{eq_angular_distribution})--(\ref{eq_linear_polarization}) represent the most general form of the angular distribution and linear polarization of photons emitted in the radiative transition between two well--defined hyperfine states $\ketm{\alpha_i F_i}$ and $\ketm{\alpha_f F_f}$. However, such \textit{individual} hyperfine transitions can hardly be resolved experimentally due to the restricted resolution of the photon detectors as well as the limitations in populating exclusively the excited state $\ketm{\alpha_i F_i}$. Therefore, only some superposition of the hyperfine components of a particular fine--structure line $\ketm{\alpha_i J_i} \to \ketm{\alpha_f J_f}$ can be measured at any given time. If we further assume that the energy splitting between the hyperfine levels $\ketm{\alpha (I J_{i,f}) F_{i,f}}$, $F_{i,f} = |I-J_{i,f}|, ... \, I+J_{i,f}$ in the initial and final ionic states is larger than that their natural widths, the linear polarization of such a superposition can be written as:
\begin{eqnarray}
   \label{eq_linear_polarization_superposition}
   P^{\rm sup}(\theta) = {\sum\limits_{F_i F_f}  N_{if} \, W_{if}(\theta) \, P_{if}(\theta) }/
   {\sum\limits_{F_i F_f} N_{if} \, W_{if}(\theta)} \, ,
\end{eqnarray}
where the $N_{if}$ are certain weight factors that describe the contribution of individual hyperfine transitions to the overall $\ketm{\alpha_i J_i} \to \ketm{\alpha_f J_f}$ fine--structure transition, and where $W_{if}(\theta)$ and $P_{if}(\theta)$ refer to Eqs.~(\ref{eq_angular_distribution}) and (\ref{eq_linear_polarization}), respectively. If, in addition, the radiative lifetimes of all excited levels are so short that virtually all the photons are counted by the detector(s), the weights $N_{if}$ are just given by the relative population of the (upper) levels $\ketm{\alpha_i F_i}$. In the following, we shall assume a population of these levels is due to electron capture and write $N_{if} \,=\, \sigma^{\rm REC}(\alpha_i F_i)/\sum_{F_i} \sigma^{\rm REC}(\alpha_i F_i)$, where $\sigma^{\rm REC}(\alpha_i F_i)$ is the total REC cross section as discussed in detail previously \cite{SuJ06, FrS05}.

%
%
%
%
\section{Evaluation of transition amplitudes}

As seen from Eqs.~(\ref{eq_angular_distribution})--(\ref{eq_linear_polarization_superposition}), the evaluation of the linear polarization of the individual hyperfine lines as well as of the average of these lines can be traced back to the calculation of capture cross sections and the reduced matrix elements $\rmem{\alpha_i F_i}{H_\gamma(p L)}{\alpha_f F_f}$. To compute these (reduced) amplitudes, one has first to generate \asr{the bound--state hyperfine wavefunctions $\ketm{\alpha F M_F}$ which are solutions of the Dirac equation:}
\begin{equation}
   \label{eq_hamiltonian}
   \left( \hat{H}_0 + \hat{H}_{\rm hf} \right) \ketm{\alpha F M_F} = E_{\alpha F} \ketm{\alpha F M_F} \, .
\end{equation}
\asr{Here, $\hat{H}_0$ is the ``unperturbed'' electronic Hamiltonian, and the operator $\hat{H}_{\rm hf}$ describes hyperfine electron--nuclear interactions. If one restricts the analysis to only include the dominant magnetic--dipole interaction,}
\begin{eqnarray}
   \label{eq_hyperfine_hamiltonian}
   \hat{H}_{\rm hf} &=&  M^{(n)}_1 \, T^{(e)}_1 \,
\end{eqnarray}
\asr{occurs as a product of (two) first--rank spherical tensor operators that operates either on the electronic ($T^{(e)}_1$) or nuclear ($M^{(n)}_1$) coordinates \cite{ChC85,Joh11}. Such a form of the interaction operator suggests that solutions of Eq.~(\ref{eq_hamiltonian}) can be formed as linear combinations of the corresponding atomic $\ketm{\beta J M_J}$ and nuclear $\ketm{I M_I}$ states:}
\begin{eqnarray}
   \label{eq_F_wavefunction}
   \ketm{\alpha F M_F} &=& \sum\limits_{\beta J} C^F_{\beta J} \ketm{\beta J I: F M_F} \nonumber \\
   &\equiv& \sum\limits_{\beta J} C^F_{\beta J} \sum\limits_{M_I M_J} \sprm{I M_I \, J M_J}{F M_F} \nonumber \\
   &\times&  \ketm{I M_I} \ketm{\beta J M_J} \, .
\end{eqnarray}
\asr{By inserting this expansion into Eq.~(\ref{eq_hamiltonian}) and performing some simple manipulations, we find that the coefficients $C^F_{\beta J}$ satisfy the secular equation:}
\begin{eqnarray}
   \label{eq_secular_equation}
   E_{F} C^F_{\beta J} &=& \sum\limits_{\beta' J'} \Bigg[ E_{\beta J} \delta_{\beta \beta'} \delta_{J J'} \nonumber \\
   && \hspace*{-2cm} + \mem{\beta J I: F M_F}{\hat{H}_{\rm hf}}{\beta' J' I: F M_F} \Bigg] C^F_{\beta' J'} \, .
\end{eqnarray}
\asr{Solution of this eigenproblem requires the knowledge of both, the unperturbned energy $E_{\beta J}$ of the electronic state $\ketm{\beta J}$ and the matrix element of the hyperfine interaction operator. After some standard angular momentum algebra the latter can be written as:}
\begin{eqnarray}
   \label{eq_hf_matrix_element}
   \mem{\beta J I: F}{\hat{H}_{\rm hf}}{\beta' J' I: F} &=& (-1)^{I + J + F} \nonumber \\
   && \hspace*{-6cm} \times \sixjm{I}{J}{F}{J'}{I}{k} \, \rmem{\beta J}{T^{(e)}_1}{\beta' J'} \rmem{I}{M^{(n)}_1}{I} \, ,
\end{eqnarray}
\asr{where $\rmem{}{T^{(e)}}{}$ and $\rmem{}{M^{(n)}}{}$ are the reduced electronic and nuclear matrix elements. While the accurate evaluation of the electronic amplitude requires detailed atomic structure calculations, as discussed in Refs.~\cite{InP89,JoC97,AbG95,Joh11,Fri12}, the nuclear term is determined geometrically by:}
\begin{eqnarray}
   \label{eq_nuclear_matrix_element}
   \rmem{I}{M^{(n)}_1}{I} &=& \mu_I \sqrt{(2I + 1)(I + 1)/I} \, ,
\end{eqnarray}
\asr{where $\mu_I$ is the nuclear magnetic dipole moment.}

\asr{As seen from Eqs.~(\ref{eq_hf_matrix_element})--(\ref{eq_nuclear_matrix_element}), the Hamiltonian matrix in the right--hand side of expression (\ref{eq_secular_equation}) and, hence, the expansion coefficients $C^F_{\beta J}$ depend in the nuclear moment $\mu_I$. Via the coefficients $C^F_{\beta J}$, moreover, the $\mu_I$--dependence enters also the amplitudes $\rmem{\alpha_i F_i}{H_\gamma(p L)}{\alpha_f F_f}$ that characterize transitions between the hyperfine levels. By making use of the expansion (\ref{eq_F_wavefunction}) these amplitudes can be obtained as:}
\begin{eqnarray}
   \label{eq_F_reduced_matrix_element}
   \rmem{\alpha_i F_i}{H_\gamma(p L)}{\alpha_f F_f} & = &
   \sum\limits_{\beta_i J_i} \sum\limits_{\beta_f J_f} \, C^{F_i}_{\beta_i J_i} \, C^{F_f}_{\beta_f J_f}
   \nonumber \\[0.15cm]
   && \hspace*{-3.5cm} \times (-1)^{J_i + I + F_f + L} \, [F_i, F_f]^{1/2} \,
                       \sixjm{F_i}{F_f}{L}{J_f}{J_i}{I} \nonumber \\[0.15cm]
   && \hspace*{-3.5cm} \times \rmem{\beta_i J_i}{H_\gamma(p L)}{\beta_f J_f} \, ,
\end{eqnarray}
where the reduced matrix element in the last line, $\rmem{\beta_i J_i}{H_\gamma(p L)}{\beta_f J_f}$, describes the usual fine--structure radiative transition. To obtain this \textit{purely electronic} amplitude, we employed the \textsc{Grasp92} \cite{grasp:96} and
\textsc{Ratip} programmes \cite{Fri12,Fri:01} that implement the multiconfiguration Dirac--Fock approach for computing the wave functions as well as different transition and ionization properties of atoms and ions.

%
%
%
%
\begin{figure}[t]
\begin{center}
\includegraphics[width=8.5cm]{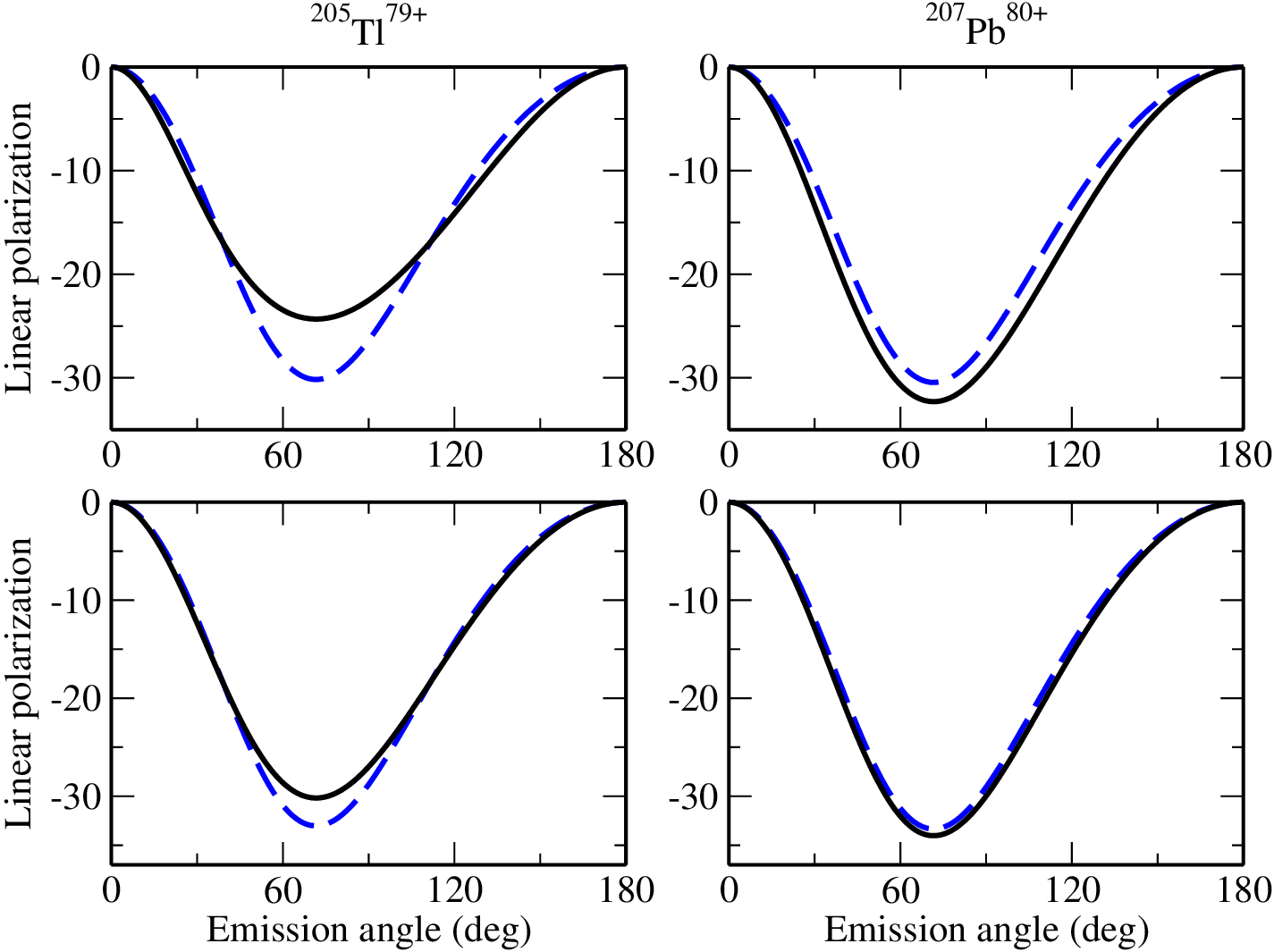}
\end{center}
\vspace*{-0.05cm}
\caption{(Color online) Degree of linear polarization of the $F_i = 3/2 \to F_f = 1/2$ hyperfine--resolved (top panel) and the averaged $1s 2p_{3/2} \;\, ^3P_2 \to 1s^2 \;\, ^1S_0$ fine--structure (bottom panel) transitions following the REC into the $1s 2p_{3/2} \;\, ^3P_2, \, F_i = 3/2, 5/2$ hyperfine levels \asr{and the cascade feeding from the high--lying states} of finally helium--like $^{205}_{~81}$Tl$^{79+}$ and $^{207}_{~82}$Pb$^{80+}$ ions. Calculations are performed within the magnetic--quadrupole approximation only (dashed lines) and by taking the $E1$--$M2$ mixing into account (solid lines). \asr{Results are presented in the laboratory frame in which the ions move with the projectile energy $T_p = 50$~MeV/u.}
\label{fig1}}
\end{figure}

%
%
%
%
%
%
%
\section{Results and discussions}

Equations~(\ref{eq_linear_polarization})--(\ref{eq_linear_polarization_superposition}) can be applied to any many--electron ion, independent of its nuclear spin and electronic shell structure. Below we shall analyze the linear polarization of the $1s 2p_{3/2} \;\, ^3P_2 \to 1s^2 \;\, ^1S_0$ line
following the REC into initially hydrogen--like ions with $I \ne 0$. Because of the hyperfine interaction, which \textit{mixes} the excited $1s 2p \;\, ^3P_2 $ state of such ions with the short--lived $1s 2p \;\, ^{1,3}P_{1}$ levels, the $1s 2p \;\,  ^3P_2 \to 1s 2p \;\, {}^1S_0$ decay can proceed also via the hyperfine--induced electric--dipole $E1$ decay in addition to the leading magnetic--quadrupole $M2$ transition \cite{BeI06,GoM74}. It is the interference between the $E1$ and $M2$ amplitudes that affects the polarization properties of the characteristic K$\alpha_1$ emission and which can be utilized as a tool for probing the nuclear spins and dipole magnetic moments.

%
%
%
%
\begin{figure}[t]
\begin{center}
\includegraphics[width=8.5cm]{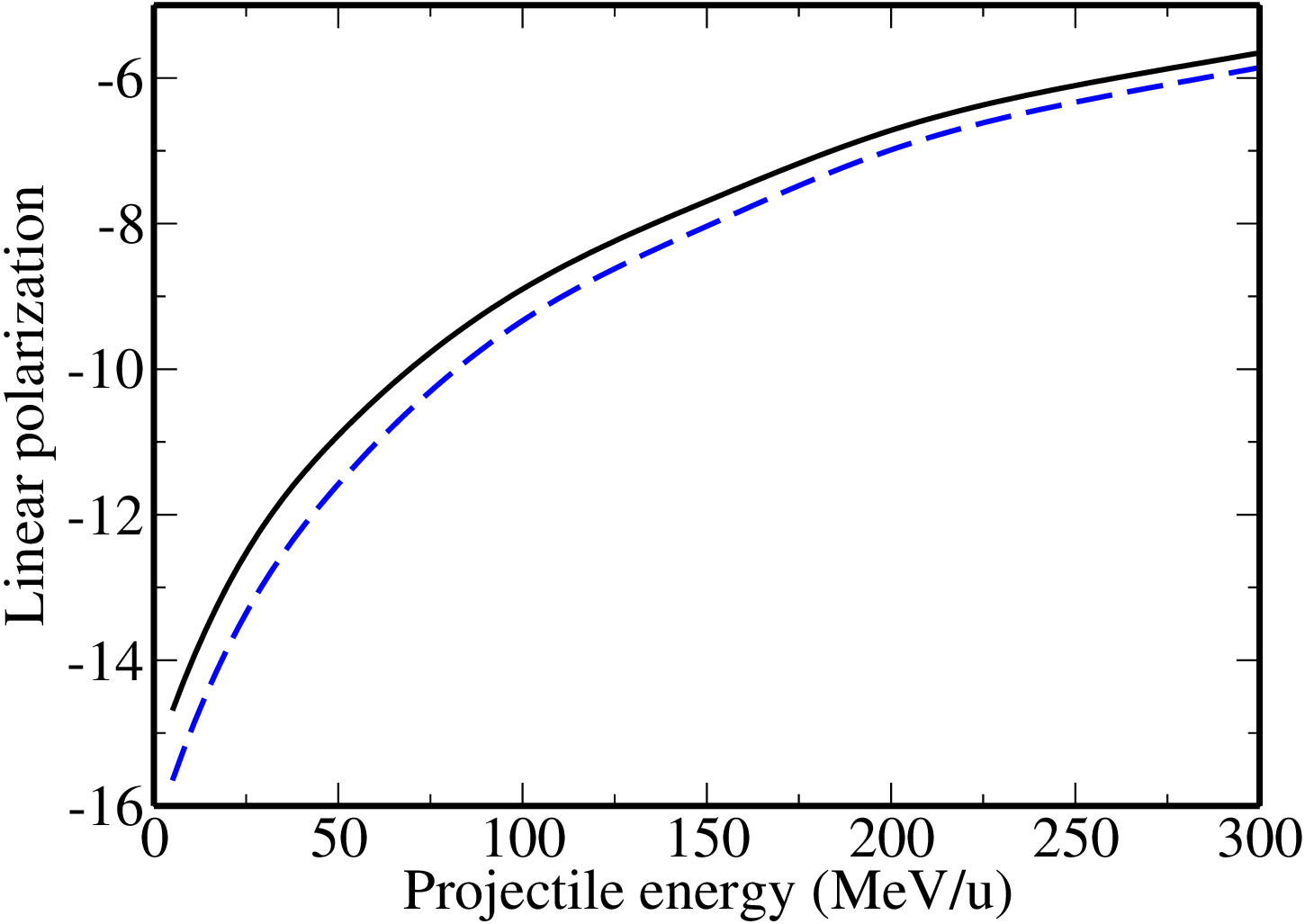}
\end{center}
\vspace*{-0.05cm}
\caption{(Color online) Degree of linear polarization of the (averaged) K$\alpha_1$ emission from helium--like thallium $^{205}_{~81}$Tl$^{79+}$ ions following the radiative electron capture into $1s 2p_{3/2} \;\, ^{1,3}P_{1,2}$ levels \asr{and the cascade feeding from the high--lying states}. Calculations within the magnetic--quadrupole approximation only (dashed line) are compared with a full account of the $E1$--$M2$ mixing (solid line). Results are presented for the emission angle $\theta = 90^\circ$ in the ion rest frame.
\label{fig2}}
\end{figure}

Owing to the angular momentum selection rules not all individual hyperfine transitions, that give rise to the overall $1s 2p \;\,  ^3P_2 \to 1s 2p \;\, {}^1S_0$ line, are influenced by the $E1$--$M2$ mixing of the transition amplitudes. For ions with nuclear spin $I = 1/2$, for example, the
hyperfine--induced $E1$ decay channel is allowed only for the $F_i = 3/2 \to F_f = 1/2$ transition. Apart from the alignment of the initial ionic state, therefore, the linear polarization of this particular transition is thus affected by the ratio of the transition amplitudes in the structure functions as:
\begin{eqnarray}
   \label{eq_structure fumctions_F}
   f_2 &\approx& - \frac{1}{2} \, \left( 1 - 2 \sqrt{3} \, \frac{a_{E1}}{a_{M2}} \right) \, , \nonumber \\
   g_2 &\approx& \frac{1}{2} \sqrt{\frac{3}{2}} \, \left( 1 + \frac{2}{\sqrt{3}} \, \frac{a_{E1}}{a_{M2}} \right) \, ,
\end{eqnarray}
where use is made of the short--hand notation
$a_{E1,M2} = \langle 1s^2 \;\, {}^1S_0, \, F_i = 1/2 \vert\vert H_\gamma(E1, M2) \vert\vert 1s2p \;\, {}^3P_2, \, $ $ F_f=3/2 \rangle$
in order to denote the corresponding (reduced) transition amplitudes. As seen from Eq.~(\ref{eq_F_reduced_matrix_element}), moreover, these amplitudes are proportional to the expansion coefficients $C^{F_f}_{\beta_f J_f}$ that describe the hyperfine--induced mixing between the $1s2p \;\, {}^3P_2$ and $^{1,3}P_{1}$ levels. For example, the matrix element $a_{E1}$ of the induced $E1$ decay reads as:
\begin{eqnarray}
   \label{eq_hfq_matrix_element}
   a_{E1} &=& \sqrt{\frac{4}{3}}
   \Bigg( C^{3/2}_{1s2p \;\, {}^1P_1} \rmem{1s2p \, {}^1P_1}{H_\gamma(E1)}{1s^2 \;\, {}^1S_0}
   \nonumber \\
   &+& C^{3/2}_{1s2p \;\, {} ^3P_1}  \rmem{1s2p \, {}^3P_1}{H_\gamma(E1)}{1s^2 \;\, {}^1S_0}  \Bigg) \, ,
\end{eqnarray}
and enables one to extract information about the mixing coefficients $C^{F_f}_{\beta_f J_f}$ and, hence, about the nuclear properties. For the known nuclear spin $I$, therefore, a polarization--resolved measurement of the $E1$--$M2$ interference on the $F_i = 3/2 \to F_f = 1/2$ hyperfine transition might give rise to a (more precise) determination of the magnetic dipole moment $\mu_I$. This is illustrated in the top panel of Fig.~\ref{fig1}, where the linear polarization of the $F_i = 3/2 \to F_f = 1/2$ transition is displayed as function of the photon emission angle $\theta$, and for the two, nuclear spin--1/2 helium--like ions, $^{205}_{~81}$Tl$^{79+}$ ($I = 1/2$, $\mu_I = 1.638\,\mu_N$) and $^{209}_{~82}$Pb$^{80+}$ ($I = 1/2$, $\mu_I = 0.593\,\mu_N$), moving with the projectile energy $T_p =50$~MeV/u. Here calculations have been made not only by taking a full account
of the hyperfine--induced $E1$ decay (solid line) but also for an non--quenched $M2$ transition (dashed lines). As seen from the figure, the admixture due to the $E1$ channel can remarkably modify the linear polarization (\ref{eq_linear_polarization}) for all angles. This mixing effect is most pronounced for large angles of the photon emission, $\theta \approx 70^\circ$ in the laboratory frame, for which the absolute value of the polarization is maximal. At these angles and for helium-like $^{205}_{~81}$Tl$^{79+}$ ions, the (absolute value of) degree $P_{if}$ reduces by about 20~\%{} from $-0.30$ to $-0.24$ if, apart from the leading magnetic--quadrupole $M2$ decay, the hyperfine--induced $E1$ amplitude is taken into account. For helium-like $^{209}_{~82}$Pb$^{80+}$ ions, in contrast, the shift in the polarization of the $F_i = 3/2 \to F_f = 1/2$ hyperfine transition does not exceed 5~\%{} at any angle because of the (much) smaller magnetic moment $\mu_I(^{209}_{~82}$Pb$) = 0.593\,\mu_N$.

Due to experimental limitations, however, it is hardly possible to directly compare the calculated polarization of the individual $F_i = 3/2 \to F_f = 1/2$ transition with measured data because of the relatively small ($\sim$ 2 eV) energy separation between the $F_i = 3/2$ and $F_i = 5/2$ levels. For this reason, only the superposition of the K$\alpha_1$ emission from the $F_i = 3/2$ and $5/2$ levels can be recorded experimentally, following the REC into initially hydrogen--like ions. The linear polarization of such an incoherently averaged $1s2p \;\, {}^3P_2 \to 1s^2 \;\, {}^1S_0$ fine--structure transition is shown in the bottom panel of Fig.~\ref{fig1}. As before, the calculations were performed within the magnetic--quadrupole approximation and by taking into account the interference between the $E1$ and $M2$ amplitudes for spin--1/2 nuclei. Although the $F_i =5/2$ level typically contributes with a larger weight than the $F_i =3/2$ one and can decay \textit{only} via a $M2$ quadrupole transition, the hyperfine--averaged $1s2p \;\, {}^3P_2 \to 1s^2 \;\, {}^1S_0$ fine--structure transition can still be clearly affected by the
$E1$--$M2$ multipole mixing. For instance, this mixing results in a reduction of the linear polarization $P^{({}^3P_2 \:\to\: {}^1S_0)}(\theta)$ by about 11~\% for $^{205}_{~81}$Tl$^{79+}$ ions, while the effect appears negligible for $^{209}_{~82}$Pb$^{80+}$ ions for the same reasons as above.

Accurate polarization measurements of the individual $1s2p \;\, {}^3P_2 \to 1s^2 \;\, {}^1S_0$ fine--structure transition are feasible today by using crystal spectrometers \cite{Hen90,Bei96}. The performance of such crystal polarimeters is however restricted to x--rays with energies $\lesssim$~10~keV. The linear polarization of more energetic photons is then usually observed by means of solid--state Compton scatterers. Due to the insufficient energy resolution of the latter devices, the $1s 2p \;\, ^3P_2 \to 1s^2 \;\, ^1S_0$ line in medium-- and high--$Z$ ions can usually also \textit{not} be resolved from the $1s 2p \;\, ^1P_1 \to 1s^2 \;\, ^1S_0$ transition. In order to better understand how the superposition of these two fine-structure components is affected by the $E1$--$M2$ mixing in the $F_i = 3/2 \,\to\, F_f = 1/2$ hyperfine channel from above, Fig.~\ref{fig2} shows the linear polarization \asr{$P^{\,{\rm K}\alpha_1}(T_p)$ of the K$\alpha_1$ ($1s2p \;\, {}^{1,3}P_{1,2} \to 1s^2 \;\, {}^1S_0$) overall emission from the helium--like $^{205}_{~81}$Tl$^{79+}$ ions as function of the projectile energy $T_p$} and for an observation angle $\theta = 90^\circ$ in the ion rest frame. Despite the averaging over both, the hyperfine-- and fine--structure components of the K$\alpha_1$ line, the interference between the $M2$ and hyperfine--induced $E1$ amplitudes of the $1s2p \;\, {}^3P_2, \, F_i = 3/2 \,\to\, 1s^2 \;\, {}^1S_0, \, F_f =1/2$ hyperfine transition is still
well seen over the entire range of projectile energies. The effect is most pronounced at low collision energies, say \asr{$T_p \lesssim$ 10 MeV/u}, for which the (absolute value of) degree of polarization is maximal. At these projectile energies, the $E1$--$M2$ mixing leads to the reduction of (the absolute value of) $P^{\,{\rm K}\alpha_1}$ from about -15.8~\%{} to -14.6~\%{}.

\begin{figure}[t]
\begin{center}
\includegraphics[width=8cm]{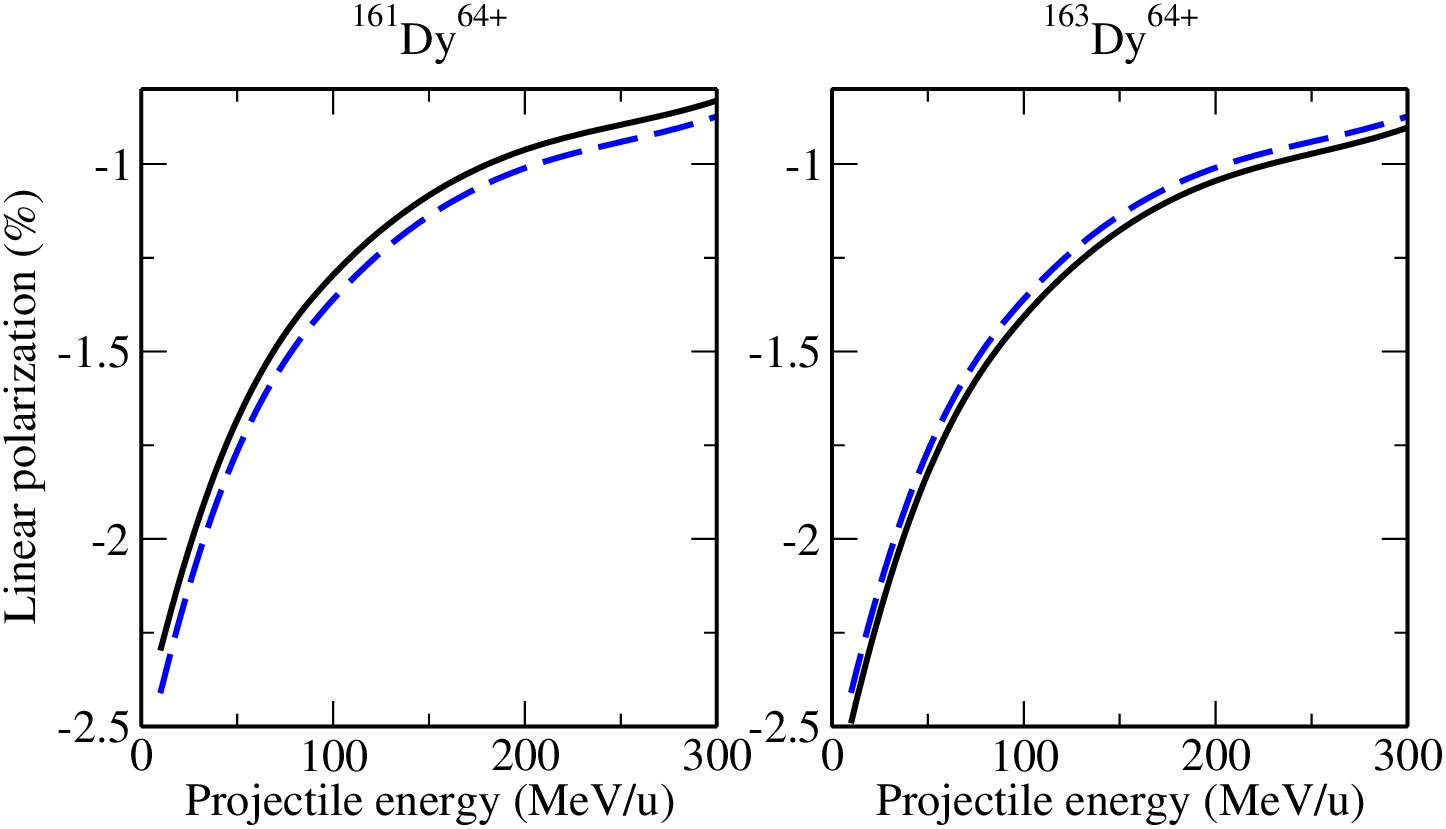}
\end{center}
\caption{(Color online) Same as Figure~\ref{fig2} but for helium-like $^{161}_{~66}$Dy$^{64+}$ (left panel) and
$^{163}_{~66}$Dy$^{64+}$ ions (right panel).
\label{fig3}}
\end{figure}

The results predicted in Figures~\ref{fig1} and \ref{fig2} clearly indicate that the linear polarization of the characteristic K$\alpha_1$ photons may provide information about the (absolute value of the) nuclear magnetic moments, if measured accurately by means of available x--ray polarimeters. Moreover, the degree of polarization, $P(\theta, T_p)$, appears sensitive also to the \textit{sign} of the magnetic dipole moment, $\mu_I$. In order to illustrate this sensitivity, Fig.~\ref{fig3} displays the linear polarization of the K$\alpha_1$ emission for helium--like $^{161}_{~66}$Dy$^{64+}$ (left panel) and $^{163}_{~66}$Dy$^{64+}$ ions (right panel). Although both dysprosium isotopes have the same nuclear spin, $I = 5/2$, and a rather similar modulus of the dipole moment, $\mu_I(^{161}_{~66}{\rm Dy}) = -0.4804\,\mu_N$ and $\mu_I(^{163}_{~66}{\rm Dy}) = +0.6726\,\mu_N$, these values differ in their sign. As seen from Fig.~\ref{fig3}, the sign change results in a different shift of the (degree of) linear polarization $P^{{\rm K}\alpha_1}$ towards \textit{lower} or \textit{higher} values. While the $E1$--$M2$ interference depolarizes the K$\alpha_1$ emission by about 4~\%{} for $^{161}$Dy$^{64+}$ ions, the degree of polarization $P^{\,{\rm K}\alpha_1}$ is enhanced by 3~\%{} for the $^{163}$Dy$^{64+}$ isotope. As usual,
the largest multipole--mixing effect can be observed at rather low collision energies for which the radiative electron capture results in a strong alignment of the excited ionic levels and, hence, in a large degree of polarization for the K$\alpha_1$ line, independent of the spin and magnetic
moment of the nuclei.

%
%
%
%

%
%
%
%
\section{Conclusions}

In summary, we have studied the polarization of the characteristic radiation of few--electron ions with \textit{non-zero} nuclear spins. Special emphasis was given to the influence of hyperfine--induced transitions and their ``multipole admixture'' in addition to the dominant decay modes allowed for nuclei with zero spin. For the $1s 2p \;\, ^3P_2 \to 1s^2 \;\, ^1S_0$ K$\alpha_1$ emission of helium--like ions, for example, the hyperfine--induced electric--dipole decay of the $1s 2p \;\, ^3P_2, \, F$ levels causes a $E1$--$M2$ mixing that strongly influences the linear polarization of the line. The analysis of this multipole mixing has been performed within the density matrix theory, and detailed computations were made with multi--configuration Dirac--Fock wave functions for selected elements and isotopes along the helium isoelectronic sequence. From these computations, it is shown that the linear polarization of the K$\alpha_1$ emission depends---more or less---sensitively on the magnetic dipole moment $\mu_I$ of the isotopes. Therefore, a polarization--resolved analysis of the characteristic emission from helium--like ions may serve also as a new tool for studying the nuclear parameters, such as the spin and magnetic moments, of different isotopes. Since simple few--electron systems are considered here, which is essential for reliable atomic calculations, these measurements may provide complementary information to conventional spectroscopy studies.

In order to explore the potential of the x--ray polarimetry for the analysis of stable or radio--active isotopes at accelerator facilities, computations have been performed especially for the K$\alpha_1$ emission of helium--like ions following the radiative electron capture into the
$1s 2p \; ^3P_1$ and $^1P_1$ levels of dysprosium, thallium and lead ions. Since the fine--structure of these excited levels cannot be resolved so easily by the present--day detectors, we have estimated also the properties of incoherent superpositions of the $1s2p \;\, ^3P_2 \to 1s^2 \,\, ^1S_0$ and $1s2p \;\, ^1P_1 \to 1s^2 \, ^1S_0$ fine--structure transitions. The linear polarization of this super--imposed K$\alpha_1$ line was shown
to be still sensitive to the $E1$--$M2$ mixing and, hence, suitable for investigating nuclear properties. Such investigations can be addressed {\it now} at heavy--ion storage rings, like, e.g., the experimental storage ring ESR at GSI, Darmstadt, where the in--flight separated beams of highly--charged radionuclides of all chemical elements up to Uranium can be stored for envisioned experiments \cite{FrG08,LiB11}. With the planned CRYRING@ESR facility, moreover, access to stored beams of highly--charged (stable and radioactive) ions at low energies $<\,10$~MeV/u will become available routinely, for which the predicted $E1$--$M2$ mixing effect can be observed most easily.

%
%
%
%
%
%


A.S.~acknowledges support from the Helmholtz Ge\-meinschaft (Nachwuchsgruppe VH--NG-421). T.~S. and Y.~L. thank Helmholtz Ge\-meinschaft for support within Helmholtz--CAS Joint Research Group (HCJRG--108).

%
%
%
%

\end{document}